\documentclass[12pt,thmsa]{article}
\usepackage{amssymb}
\usepackage{graphicx}

\input{tcilatex}
\begin{document}

\begin{center}

{\Large Entanglement sudden birth of two trapped ions interacting
with a time-dependent laser field}

M. Abdel-Aty$^1$ and T. Yu$^2$

{\small $^1$Mathematics Department, College of Science, Bahrain
University, 32038 Kingdom of Bahrain}

{\small $^2$Rochester Theory Center and Department of Physics and
Astronomy, University of Rochester, New York 14627, USA}

\end{center}

Journal of Physics B: At. Mol. Opt. Phys. 41 (2008) 235503

We explore and develop the mathematics of the two multi-level ions.
In particular, we describe some new features of quantum entanglement
in two three-level trapped ions confined in a one-dimensional
harmonic potential, allowing the instantaneous position of the
center-of-mass motion of the ions to be explicitly time-dependent.
By solving the exact dynamics of the system, we show how
survivability of the quantum entanglement is determined by  a
specific choice of the initial state settings.




\section{Introduction}
The ability to laser cool ions within the vicinity of the motional
ground state is a key factor in the realization of efficient quantum
communication \cite{win98}. Laser-cooled ions confined in an
electromagnetic trap are good candidates for various quantum
information processing processes such as quantum control and quantum
computing \cite{cirac95}. Various interactions models and higher
order non-linear models can be implemented in this system by simply
choosing appropriate applied laser tunings \cite{blo92,vog95}. For
more than one ion, as required in any realistic quantum logic
schemes, individual addressing imposes small trap frequencies,
whereas sideband cooling imposes high trap frequencies \cite{mor99}.
Experimentally, the sideband cooling of two ions to the ground state
has been achieved in a Paul trap that operates in the Lamb-Dicke
limit \cite{kin98}. Recent advances in the dynamics of trapped ions
\cite{lei03}) have demonstrated that a macroscopic observer can
effectively control dynamics as well as perform a complete
measurement of states of microscopic quantum systems.

On the other hand, entanglement is the quintessential property of
quantum mechanics that sets it apart from any classical physical
theory, and it is essential to quantify it in order to assess the
performance of applications of quantum information processing
\cite{hil97}. With the reliance in the processing of quantum
information on a cold trapped ion, a long-living entanglement in the
ion-field interaction with pair cat states has been observed
\cite{abd06}. Also, experimental preparation and measurement of the
motional state of a trapped ion, which has been initially laser
cooled to the zero-point of motion, has been reported in
\cite{lei97}. It is well known that the loss of entanglement cannot
be restored by local operations and classical communications, which
is one of the main obstacles to build up a quantum computer
\cite{ben96}. Therefore it becomes an important subject
to study entanglement dynamics of qubits \cite%
{yu06,yon06,pin06,yu02,yu03,yu04}. Quite recently, it has been shown
that entanglement of two-qubit systems can be terminated abruptly in
a finite time in the presence of either quantum or classical noises
\cite{yu06,yu04}. This non-smooth finite-time decay has been
referred as entanglement sudden death (ESD)
\cite{yu06,yu06prl,Yu-Eberly07}. At least two experimental
verifications of ESD have been reported so far \cite{exp1,exp2}.
Interestingly, as the entanglement can suddenly disappear, it can be
suddenly generated, a process called entanglement sudden birth
\cite{lop08}.

These recent results together with the fundamental interest of the
subject, are the motivation for the present work: we study here the
entanglement evolution of two three-level trapped ions interacting
with a laser field taking into account the time-dependent modulated
function. We focus our attention on the entanglement decay and
generation due to the presence of a time-dependent modulated
function in the two interacting ions and intrinsic decoherence. We
show that entanglement of two ions in the Lamb-Dicke regime presents
some novel features with respect to single-trapped ion. In
particular, we show how sudden feature in the entanglement(sudden
birth and sudden death of entanglement) occurs for the two-ion
system coupled to a laser field.

The paper is organized as follows. In section 2, we introduce the
model and define the different parameters which will be used
throughout the paper. We devote section 3 to an alternative approach
to find an exact dynamics of the coupled system in the presence of
the time-dependent modulated function. Section 4 is devoted to study
some examples of entanglement evolutions which lead to different
types of decay for entangled systems, and show how different initial
state settings and time-dependent interaction affect the decay of
the entanglement. Then we conclude the paper with some remarks on
how the presented results could be generalized to multi-ions
systems.

\section{Model}

\label{model}

A useful model of controlled entanglement evolution consists of two
three-level ions that are irradiated by laser beams whose
wavevectors lie in the radial plane of the trap. This laser
configuration excites the
vibrational motion in the radial plane only ($1D$ ionic motion) \cite%
{mes03,sha02}. Therefore, the physical system on which we focus is
two three-level harmonically trapped ions with their center-of-mass
motion quantized. We denote by $a$ and $a^{\dagger }$ the
annihilation and creation operators and $\upsilon $ is the
vibrational frequency related to the center-of-mass harmonic motion
along the direction $\hat{x}.$ Without the rotating wave
approximation, the trapped ions Hamiltonian for the system of
interest may be written as
\begin{equation}
\hat{H}=\hat{H}_{cm}+\hat{H}_{ion}+\hat{H}_{int},  \label{1}
\end{equation}%
where
\begin{eqnarray}
\hat{H}_{cm} &=&\hbar \upsilon \hat{a}^{\dagger }\hat{a},\qquad  \nonumber \\
\hat{H}_{ion} &=&\sum\limits_{i=1}^{2}\sum\limits_{j=a,b,c}\hbar
\omega
_{i}S_{jj}^{(i)},  \nonumber \\
\hat{H}_{int}(t) &=&\hbar \Im (\hat{x},t)\hat{S}_{ab}^{(1)}+\hbar
\Im ^{\ast }(\hat{x},t)\hat{S}_{ba}^{(1)}+\hbar \Im
(\hat{x},t)\hat{S}_{ab}^{(2)}+\hbar
\Im ^{\ast }(\hat{x},t)\hat{S}_{ba}^{(2)}  \nonumber \\
&&+\hbar \Im (\hat{x},t)\hat{S}_{ac}^{(1)}+\hbar \Im ^{\ast }(\hat{x},t)\hat{%
S}_{ca}^{(1)}+\hbar \Im (\hat{x},t)\hat{S}_{ac}^{(2)}+\hbar \Im ^{\ast }(%
\hat{x},t)\hat{S}_{ca}^{(2)}.  \label{ham}
\end{eqnarray}%
We denote by $\hat{S}_{lm}^{(i)}$ the atomic flip operator for the $%
|m\rangle _{i}\rightarrow |l\rangle _{i}$ transition between the two
electronic states, where $\hat{S}_{lm}^{(i)}=|l\rangle _{ii}\langle
m|,(l,m=a,b,c)$. Suppose the ions are irradiated by a laser field of
the form $\Im (\widehat{r},t)=\hbar ^{-1}\epsilon \langle i|d.\wp
|j\rangle \exp [-i(k\widehat{r}-\Omega t)],$ where $\epsilon $ is
the amplitude of the laser field with frequency $\Omega $ and
polarization vector $\wp $. The transition in the three-level ions
is characterized by the dipole moment $d$ and $k$ is the wave vector
of the laser field. Therefore if we express the center of mass
position in terms of the creation and annihilation operators of the
one-dimensional trap namely $\hat{x}=\Delta x(a^{\dagger }+a),$
where $\Delta x=(\hbar /2\upsilon m)^{1/2}=\eta /k$ is the widths of
the dimensional potential ground states, in the $x$ direction ($\eta
$ is called Lamb--Dicke parameter describing the localization of the
spatial extension of the center-of-mass), and $m$ is the mass of the
ions.

By using a special form of Baker-Hausdorff theorem \cite{blo92}, the
operator $\exp [i\eta (a^{\dagger }+a)]$ may be written as a product
of
operators i.e. $\exp (i\eta (a^{\dagger }+a))=\exp \left( \frac{\eta ^{2}}{2}%
[a^{\dagger },a]\right) \exp \left( i\eta a^{\dagger }\right) \exp
\left( i\eta a\right).$ The physical processes implied by the
various terms of the operator
\begin{equation}
\exp \left( i\eta \left( a^{\dagger }+a\right) \right) =\exp \left( \frac{%
-\eta ^{2}}{2}\right) \sum_{n=0}^{\infty }\frac{\left( i\eta \right)
^{n}a^{\dagger n}}{n!}\sum_{m=0}^{\infty }\frac{\left( i\eta \right)
^{m}a^{m}}{m!},
\end{equation}%
may be divided into three categories (i) the terms for $n>m$
correspond to an increase in energy linked with the motional state
of center of mass of the ion by ($n-m$) quanta, (ii) the terms with
$n<m$ represent destruction of ($m-n$) quanta of energy thus
reducing the amount of energy linked with the center of mass motion
and (iii) ($n=m$), represents the diagonal contributions. When we
take Lamb-Dicke limit and apply the rotating wave approximation
discarding the rapidly oscillating terms, the interaction
Hamiltonian (\ref{ham}) takes a simple form
\begin{eqnarray}
\hat{H}_{int} &=&\sum\limits_{i=1}^{2}\left( \hbar \lambda _{1}^{(i)}(t)%
\mathcal{E}(a^{\dagger }a)\hat{S}_{12}^{(i)}a^{\dagger }+\hbar
\lambda
_{1}^{(i)\ast }(t)\mathcal{E}^{\ast }(a^{\dagger }a)\hat{S}%
_{21}^{(i)}a\right)  \nonumber \\
&&+\left( \hbar \lambda _{2}^{(i)}(t)\mathcal{E}(a^{\dagger }a)\hat{S}%
_{13}^{(i)}a^{\dagger }+\hbar \lambda _{2}^{(i)\ast
}(t)\mathcal{E}^{\ast }(a^{\dagger }a)\hat{S}_{31}^{(i)}a\right) ,
\label{ham4}
\end{eqnarray}%
where $\lambda _{j}^{(i)}(t)$ is a new coupling parameter adjusted
to be time dependent. The other contributions which rapidly
oscillate with frequency $\nu $ have been disregarded. Note that in
the Lamb-Dicke regime only processes with first and second order of
$\eta $ are considered, while in the general case, the nonlinear
coupling function is derived by expanding the operator-valued mode
function as
\begin{equation}
\mathcal{E}_{k}(a^{\dagger }a)=-\frac{\epsilon }{2}\exp \left(
-\frac{\eta
^{2}}{2}\right) \sum\limits_{n=0}^{\infty }\frac{(i\eta )^{2n+k}}{n!(n+k)!}%
a^{\dagger n}a^{n}.
\end{equation}%
Since $\mathcal{E}_{k}(a^{\dagger }a)$ depends only on the quantum number $%
a^{\dagger }a$, in the basis of its eigenstates, $a^{\dagger
}a|n\rangle =n|n\rangle $, ($n=0,1,2,...$), these operators are
diagonal, with their diagonal elements $\langle
n|\mathcal{E}_{k}(a^{\dagger }a)|n\rangle $ is given by
$\mathcal{E}_{k}^{(j)}(n)=-0.5\epsilon (n+k)!)^{-1}n!L_{n}^{k}(\eta
^{2})\exp \left( -\eta ^{2}/2\right) $ where $L_{n}^{k}(\eta ^{2})$
are the associated Laguerre polynomials.\bigskip

\section{Solution}

\subsection{Master equation}

To study the time evolution of the quantum entanglement, we will
start by
finding the general solution of the two-ion system with the Hamiltonian ( %
\ref{ham4}). \ The intrinsic decoherence approach \cite{mil91}, is
based on the assumption that on sufficiently short time scales the
system does not evolve continuously under unitary evolution but
rather in a stochastic sequence of identical unitary
transformations. Under the Markov approximation \cite{mil91}, the
master equation governing the time evolution for the two-ion system
under the intrinsic decoherence is given by
\begin{equation}
\frac{d}{dt}\hat{\rho}(t)=-\frac{i}{\hbar
}[\hat{H},\hat{\rho}]-\frac{\gamma }{2\hbar
^{2}}[\hat{H},[\hat{H},\hat{\rho}]],  \label{mas}
\end{equation}%
where $\gamma $ is the intrinsic decoherence parameter. The first
term on the right-hand side of equation (\ref{mas}) generates a
coherent unitary evolution of the density matrix, while the second
term represents the decoherence effect on the system and generates
an incoherent dynamics of the coupled system. In order to obtain an
exact solution for the density operator $\hat{\rho}(t)$ of the
master equation (\ref{mas}), three auxiliary superoperators
$\widehat{J},\widehat{S}$ and $\widehat{L}$ can be introduced as
($\hbar =1)$
\begin{eqnarray}
\exp \left( \widehat{J}\tau \right) \hat{\rho}(t)
&=&\sum\limits_{k=0}^{\infty }\frac{\left( \gamma \tau \right) ^{k}}{k!}\hat{%
H}^{k}\hat{\rho}(t)\hat{H}^{k},  \nonumber \\
\exp \left( \widehat{S}\tau \right) \hat{\rho}(t) &=&\exp \left( -i\hat{H}%
\tau \right) \hat{\rho}(t)\exp \left( i\hat{H}\tau \right) , \\
\exp \left( \widehat{L}\tau \right) \hat{\rho}(t) &=&\exp \left(
-\frac{\tau
\gamma }{2}\hat{H}^{2}\right) \hat{\rho}(t)\exp \left( -\frac{\tau \gamma }{2%
}\hat{H}^{2}\right) ,  \nonumber
\end{eqnarray}%
where $\widehat{J}\hat{\rho}(t)=\gamma \hat{H}\hat{\rho}(t)\hat{H},$ $%
\widehat{S}\hat{\rho}(t)=-i[\hat{H},\hat{\rho}(t)]$ and $\widehat{L}\hat{\rho%
}(t)=-(\gamma
/2)(\hat{H}^{2}\hat{\rho}(t)+\hat{\rho}(t)\hat{H}^{2}).$ Then, it is
straightforward to write down the formal solution of the master
equation (\ref{mas}) as follows
\begin{eqnarray}
\hat{\rho}(t) &=&\exp (\widehat{J}t)\exp (\widehat{S}t)\exp (\widehat{L}t)%
\hat{\rho}(0)  \nonumber \\
&=&\sum_{k=0}^{\infty
}\hat{M}_{k}(t)\hat{\rho}(0)\hat{M}_{k}^{\dagger }(t). \label{dens}
\end{eqnarray}%
We denote by $\hat{\rho}(0)$ the density operator of the initial
state of
the system and $\hat{M}_{k}=\frac{\left( \gamma t\right) ^{k/2}}{\sqrt{k!}}%
\hat{H}^{k}\exp \left( -i\hat{H}t\right) \exp \left( -\frac{\gamma t}{2}\hat{%
H}^{2}\right) $, where the so-called Kraus operators $\hat{M}_{k}$ satisfy $%
\sum\limits_{k=0}^{\infty }\hat{M}_{k}(t)\hat{M}_{k}^{\dagger
}(t)=\hat{I}$ for all $t$.

Eq.~(\ref{dens}) can also be written as
\begin{equation}
\hat{\rho}(t)=e^{-i\hat{H}t}\exp \left( -\frac{\gamma t}{2}\hat{H}%
^{2}\right) \left\{ e^{\left( \hat{S}_{M}t\right)
}\hat{\rho}(0)\right\} \exp \left( -\frac{\gamma
t}{2}\hat{H}^{2}\right) e^{i\hat{H}t}, \label{rho1}
\end{equation}%
where, we define the superoperator $\hat{S}_{M}\hat{\rho}(0)=\gamma \hat{H}%
\hat{\rho}(0)\hat{H}$. The reduced density matrix of the ions is
given by taking the partial trace over the field system, i.e. $\rho
^{a}(t)\equiv
tr_{F}\rho (t)=\rho _{ij,lk}(t)|ij\rangle \langle lk|.$ We use the notation $%
|ij\rangle =|i_{1}\rangle \otimes |j_{2}\rangle ,$ $(i,j=a,b,c),$ where $%
|a_{1(2)}\rangle ,$ $|b_{1(2)}\rangle $ and $|c_{1(2)}\rangle $ are
the basis states of the first (second) ion and $\rho
_{ij,lk}(t)=\langle ij|\rho ^{a}(t)|lk\rangle $ corresponds the
diagonal ($ij=lk$) and off-diagonal ($ij\neq lk)$ elements of the
density matrix $\rho^a(t).$

It is worth referring here to the fact that, Eq.~(\ref{rho1})
represents the general solution of the master equation (\ref{mas})
that can be used to
study the effect of the intrinsic decoherence when the modulated function $%
\lambda _{j}^{(i)}(t)$ is taken to be time-independent. However, to
study the time-dependent case as seen below, it will be much more
convenient to use the exact wave function.

\subsection{Wave function}

At any time the state vector $\left\vert \Psi (t)\right\rangle $ in
the interaction picture can be written as
\begin{equation}
\left\vert \Psi (t)\right\rangle
=\sum\limits_{i=1}^{9}\sum\limits_{n=0}^{\infty
}A_{i}(n,t)\left\vert \psi _{i}\right\rangle .  \label{psi}
\end{equation}%
where $\left\vert \psi _{1}\right\rangle =\left\vert
n;a_{1}a_{2}\right\rangle ,\left\vert \psi _{2}\right\rangle
=\left\vert n+1;a_{1}b_{2}\right\rangle ,$ $\left\vert \psi
_{3}\right\rangle =\left\vert n+1;a_{1}c_{2}\right\rangle ,$
$\left\vert \psi _{4}\right\rangle =\left\vert
n+1;b_{1}a_{2}\right\rangle ,\left\vert \psi _{5}\right\rangle
=\left\vert n+2;b_{1}b_{2}\right\rangle ,$ $\left\vert
\psi _{6}\right\rangle =\left\vert n+2;b_{1}c_{2}\right\rangle ,$ $%
\left\vert \psi _{7}\right\rangle =\left\vert
n+1;c_{1}a_{2}\right\rangle ,$ $\left\vert \psi _{8}\right\rangle
=\left\vert n+2;c_{1}b_{2}\right\rangle ,$ $\left\vert \psi
_{9}\right\rangle =\left\vert n+2;c_{1}c_{2}\right\rangle ,$ where
$\left\vert n\right\rangle $ corresponds to the initial state of the
vibrational motion. We denote by $A_{i}(n,t)$ the probability
amplitude for the ion to be in the state $\left\vert \psi
_{i}\right\rangle $. In general,
the dynamics of the probability amplitudes $A_{i}(n,t)$ is given by the Schr%
\"{o}dinger equation,
\begin{equation}
i\frac{\partial A_{j}(n,t)}{\partial t}=\sum%
\limits_{l=1}^{9}C_{jl}A_{l}(n,t),  \label{schr}
\end{equation}%
where $C_{jk}=\left\langle \psi _{j}\right\vert
\hat{H}_{int}\left\vert \psi _{k}\right\rangle .$ These equations
are exact for any two three-level
system. In order to consider the most general case, we solve equation (\ref%
{schr}), by assuming a new variable $G(n,t)$ as $G(n,t)=\sum
\limits_{i=1}^{9}x_{i}A_{i}(n,t),$where $x_{1}=1,$\ which means that
\begin{equation}
i\frac{dG(n,t)}{dt}=\gamma (t)\beta _{1}\left(
A_{1}(n,t)+\sum\limits_{i=2}^{9}\beta _{i}^{\prime
}A_{i}(n,t)\right) , \label{amp}
\end{equation}%
where $\lambda _{j}^{(i)}(t)=\gamma (t)\lambda _{j}$ and $\beta
_{i}$ are
given by $\beta _{1}=x_{2}C_{21}+x_{3}C_{31}+x_{4}C_{41}+x_{7}C_{71},$ and $%
\beta _{i}^{\prime }=\beta _{i}/\beta _{1},$ where $\beta
_{2}=C_{11}+x_{5}C_{51}+x_{8}C_{81},$ $\beta
_{3}=C_{13}+x_{6}C_{63}+x_{9}C_{93},$ $\beta
_{4}=C_{14}+x_{5}C_{54}+x_{6}C_{64},$ $\beta _{5}=x_{2}C_{25}+x_{4}C_{45},$ $%
\beta _{6}=x_{3}C_{36}+x_{4}C_{46},$ $\beta
_{7}=C_{17}+x_{8}C_{87}+x_{9}C_{97},$ $\beta _{8}=x_{2}C_{28},$
$\beta _{9}=x_{3}C_{39}+x_{7}C_{79},$

Let us emphasize that in addition to the general form of equation
(\ref{amp} ), the present method is also suitable for any initial
conditions. It is instructive to examine the formation of a general
solution of the two three-level systems. Therefore, we use equation
(\ref{amp}) and seek $G(n,t)$ such that $i\dot{G}(n,t)=z\zeta
(t)G(n,t).$ This holds if $\beta _{1}=z$ and $\beta _{i}=x_{i}.$
Some simple algebra gives rise to an equation that has nine
eigenvalues such that the $z_{i}$ \cite{gui03} are determined.
Correspondingly, there are also nine corresponding eigenfunctions
\begin{equation}
G_{j}(n,t)=G_{j}(0)\exp \left( -iz_{j}\int_{0}^{t}\zeta (\tau )d\tau
\right) ,
\end{equation}%
where $G_{j}(n,t)=\sum\limits_{k=1}^{9}M_{jk}A_{k}(n,t),$ and $M_{ji}=%
\widehat{p}^{T}\hat{e}_{xi}+\sum\limits_{i=1}^{9}\hat{x}_{i}^{T}\hat{e}%
_{xi}. $ We denote by $\hat{e}_{xi}$ mutually orthogonal unit
vectors, given by $\hat{e}_{x_{i}}=(\delta _{i1},\delta _{i2},\delta
_{i3},\delta
_{i4},\delta _{i5},\delta _{i6},\delta _{i7},\delta _{i8},\delta _{i9}),$ $%
\delta _{ij}=1$ if $i=j$ and $\delta _{ij}=0$ if $i\neq j$ and $\widehat{p}%
=(1,1,1,1,1,1,1,1,1),$ $\hat{x}_{i}=(x_{i1},x_{i2},x_{i3},x_{i4},$ $%
x_{i5},x_{i6},$ $x_{i7},$ $x_{i8},x_{i9}).$ Then, one can express
the unperturbed state amplitude $A_{i}(n,t)$ in terms of the dressed
state amplitude $G_{j}(n,t)$ as follows
\begin{equation}
A_{i}(n,t)=\sum\limits_{j=1}^{9}M_{ij}^{-1}G_{j}(n,t)=\sum%
\limits_{j=1}^{9}M_{ij}^{-1}G_{j}(0)\exp \left(
-iz_{j}\int_{0}^{t}\zeta (\tau )d\tau \right) .
\end{equation}%
We have thus completely determined the exact solution of a two
three-level system in the presence of the time-dependent modulated
function.

\section{Entanglement evolution}

Good measures of entanglement are invariant under local unitary
operations. They are also required to be entanglement monotones,
that is, they must be non-increasing under local quantum operations
combined with classical communication
\cite{vid00,pho88,pho91,pho91a,ber03,bos01}. computable entanglement
measures for a generic multiple level system do not exist. For
two-qubit systems, concurrence can be used to compute entanglement
for both pure and mixed states \cite{woo98}. Rungta et al
\cite{run01} defined the so-called \textit{I-}concurrence in terms
of a universal-inverter which is a generalization to higher
dimensions of two spin flip operation, therefore, the concurrence in
arbitrary dimensions takes the form
\begin{equation}
I_{\psi }\left( t\right) =\sqrt{\langle \psi |S_{N_{1}}\otimes
S_{N_{1}}(|\psi \rangle \langle \psi |)|\psi \rangle }. \label{conc}
\end{equation}%
Another generalization is proposed by Audenaert et al \cite{aud01}
by defining a concurrence vector in terms of a specific set of
antilinear operators. As a complete characterization of entanglement
of a bipartite state in arbitrary dimensions may require a quantity
which, even for pure states, does not reduce to single number
\cite{mey02}, Fan \textit{et al.} defined the concept of a
concurrence hierarchy as $N-1$ invariants of a group of local
unitary for $N-$level systems \cite{fan03}.

\begin{figure}[tbph]
\begin{center}
\includegraphics[width=7cm,height=7cm]{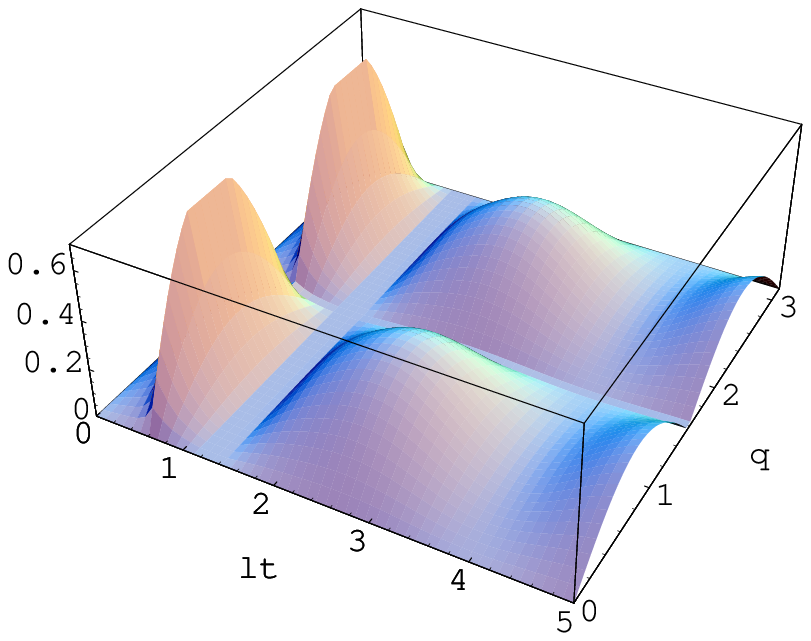} %
\includegraphics[width=6cm,height=5cm]{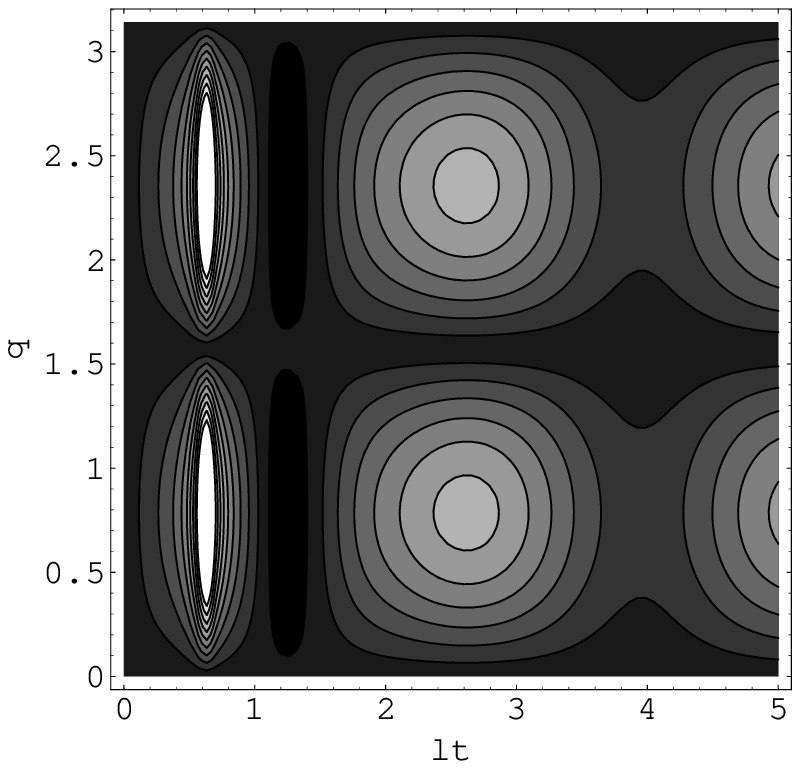}
\end{center}
\caption{The evolution of the concurrence $I_{\protect\psi }\left(
t\right) $
as a function of the scaled time $\protect\lambda _{1}t$ and $\protect\theta %
.$ The parameters are $\overline{n}=5,$ $\protect\lambda _{2}/\protect%
\lambda _{1}=0.01$, and $\protect\varphi =0.$}
\end{figure}
Now, we suppose that the initial state of the principle system is
given by
\begin{equation}
|\psi (0)\rangle =\left( \cos \theta |a_{1},b_{2}\rangle +\sin
\theta e^{i\phi }|b_{1},a_{2}\rangle \right) \otimes
\sum\limits_{n=0}^{\infty }q_{n}|n\rangle ,
\end{equation}%
where $\theta \in \lbrack 0,2\pi ],$ $\phi \in \lbrack 0,\pi ]$ and $%
q_{n}=\langle n|\psi (0)\rangle .$

In absence of intrinsic decoherence, the concurrence (\ref{conc})
will be calculated using Eq. (\ref{psi}), while to study the effect
of $\gamma $ we use equation (\ref{rho1}). In figures 1 and 2 the
dependence on $\theta $ and the scaled time $\lambda _{1}t,$ in the
Lamb-Dicke regime, of the concurrence $I_{\psi }\left( t\right) $ is
illustrated. In the typical experiments at NIST \cite{lei00},
$^{9}Be^{+}$ ions are stored in a RF Paul trap with a secular
frequency along $\widehat{x}$ of $\nu /2\pi \simeq 11.2$ MHz,
providing a spread of the ground state wave function of $\Delta
x\simeq 7$ nm, with a Lamb-Dicke parameter of $\eta \simeq 0.202$.
With these data we find $\epsilon \simeq 0.01$, so they can be
considered as small parameters. We see that the concurrence $I_{\psi
}\left( t\right) $ exhibits some peaks whose amplitude decreases as
the interaction time increases. The main consequence is that we can
select a given value of the superposition parameter $\theta $, or
rather a specific instant time to obtain strong entanglement or
disentanglement.

The entanglement sudden birth phenomenon is observed in figure 1.
Entanglement is not present at earlier times, and suddenly at some
finite time an entanglement starts to build up. Also, from figure 1,
it is clearly seen that the value of first local maximum
significantly exceeds the second local maximum when the two ions
start from a superposition state. Recall that the entanglement
attains the zero value (i.e., disentanglement) when
the trapped ions start from either $|a_{1},b_{2}\rangle $ or $%
|b_{1},a_{2}\rangle $ states, while strong entanglement occurs when
the inversion is equal to zero, i.e, the two ions start from a
maximum entangled state, such as $|\psi _{AB}(0)\rangle =\left(
|a_{1},b_{2}\rangle +|b_{1},a_{2}\rangle \right) /\sqrt{2}$. In
other words, for the initial entangled state, there are some
intervals of the interaction time where the entanglement reaches its
local maximum and drops to zero (entanglement sudden death).

\begin{figure}[tbph]
\begin{center}
\includegraphics[width=7cm,height=7cm]{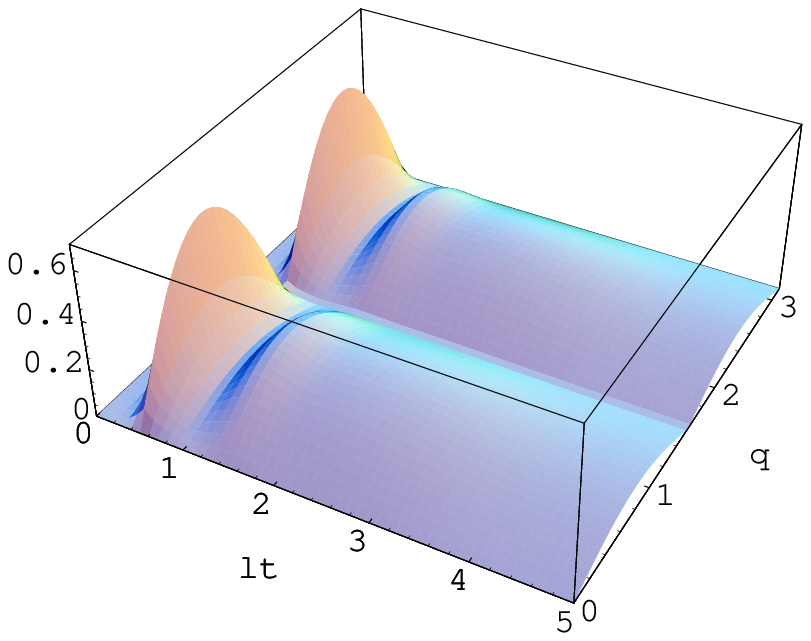} %
\includegraphics[width=6cm,height=5cm]{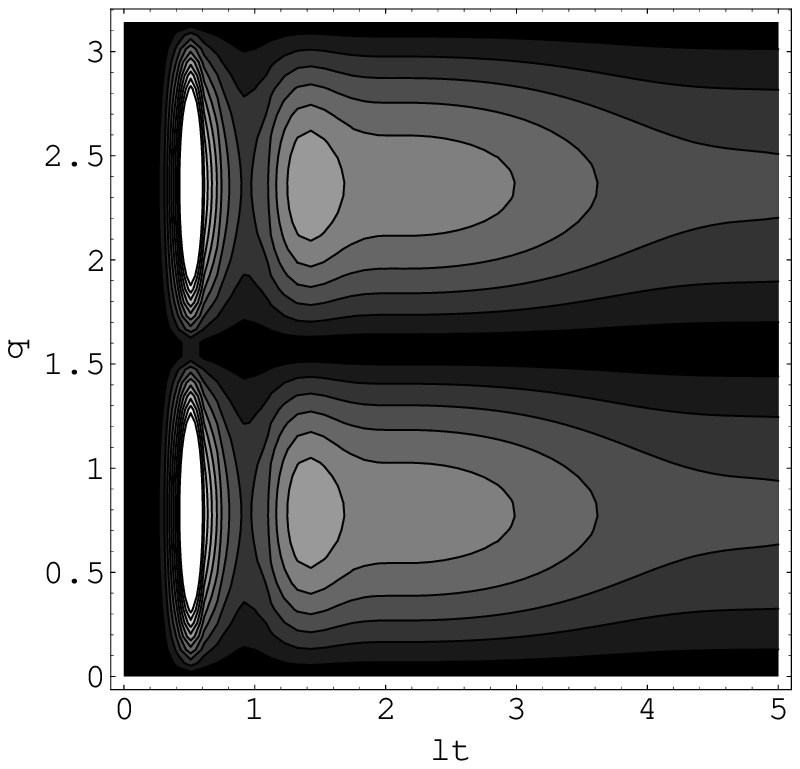}
\end{center}
\caption{The same as figure 1 but $\overline{n}=15.$}
\end{figure}

To gain more insight into the general behavior of the quantum
entanglement evolution, we plot in figure 2 the time evolution of
the generalized concurrence $I_{\psi }\left( t\right) $ for a larger
value of the field intensity, $\overline{n}$. It can be seen that
the dynamics is strongly modified where a smooth decay of the
entanglement is observed. In contrast to the dynamics with the small
$\overline{n}$, where the sudden death of entanglement-like feature
has been observed after the first local maximum of the entanglement
(see figures 1 and 2), here the entanglement smoothly decays and
vanishes as the time increased further. Note that the effect of the
superposition parameter $\theta $ on the entanglement distributions
in both figure 1 and 2 is similar and shows symmetry around $\theta
=\frac{\pi
}{2}$ while the local maxima correspond to $\theta =\frac{n\pi }{4}%
,(n=1,3,5,...)$. Obviously from these figures, if $\theta =\frac{n\pi }{2}%
,(n=0,1,2,...)$ i.e., the system starts from a separable state, the
entanglement is always zero. On the other hand, if the system starts
from an entangled state and because of the existence of further
system parameters, the entanglement decay occurs with small or large
values of $\overline{n}$. Moreover, comparison of figures 1-2 shows
that the entanglement decay takes a longer time to reach zero in the
case where large $\overline{n}$ is considered. We hence come to
understand that considering a large intensity of the initial state
of the field, can be used positively in preventing entanglement
sudden death or delaying the disentanglement of the two ions. A
physical explanation of why is the intensity of the field playing
such a role in the disentangling process of the two ions, is that
the strong field providing some sort of shielding mechanism to the
decoherence effectively induced by the trace over the motional
degree of freedom.

\begin{figure}[tbph]
\begin{center}
\includegraphics[width=7cm,height=7cm]{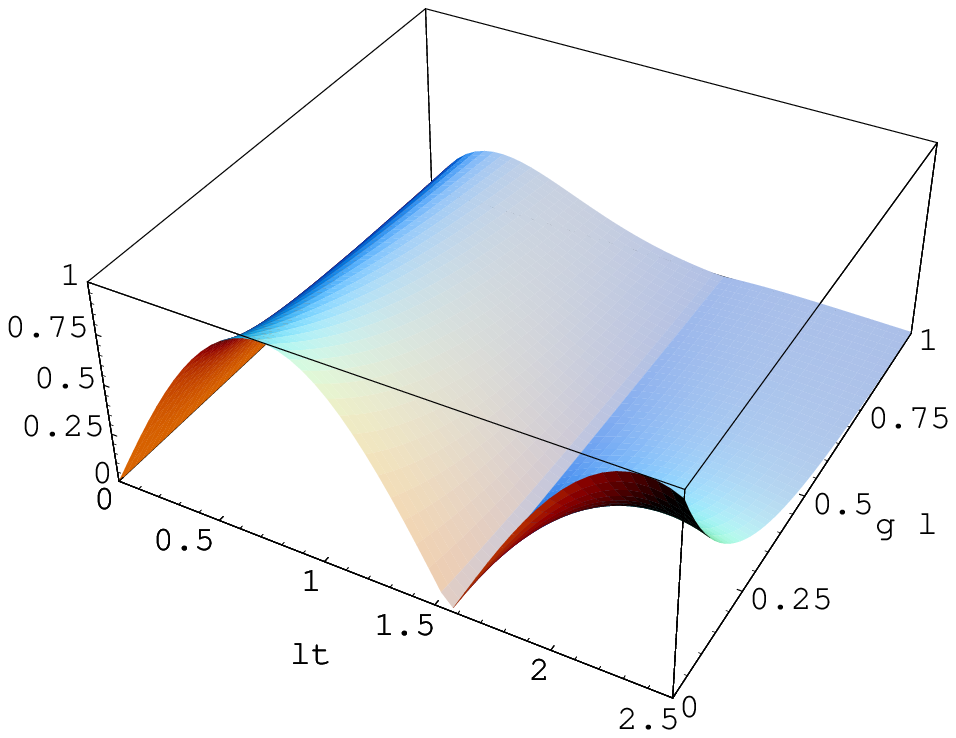} %
\includegraphics[width=6cm,height=5cm]{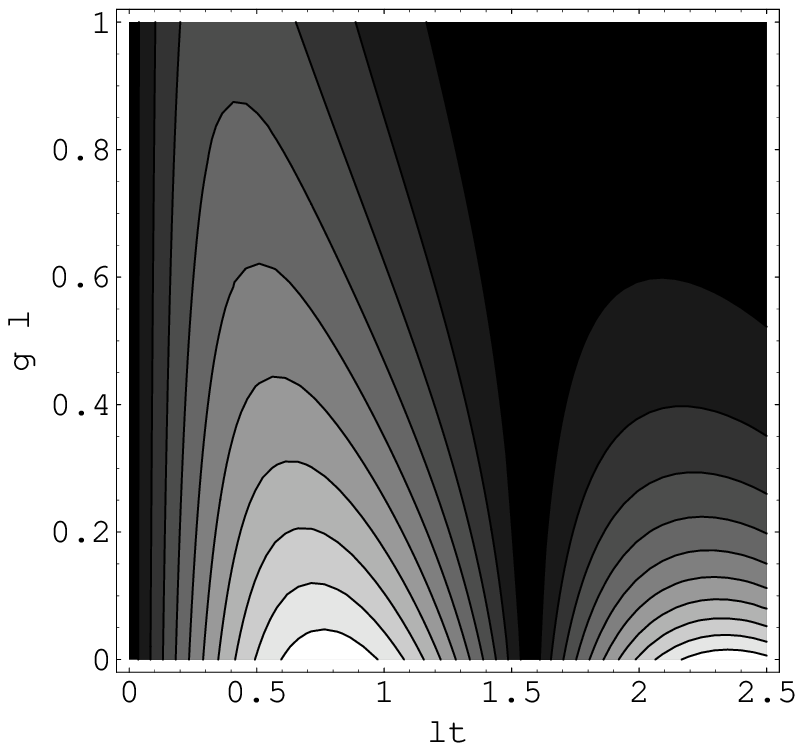}
\end{center}
\caption{The concurrence as a function of the scaled time and
decoherence parameter. The other parameters are the same as figure
1.} \label{dec}
\end{figure}

Quantum coherence and entanglement typically decay as the result of
the influence of decoherence and much effort has been directed to
extend the coherence time of systems of interest. However, it has
been shown that under particular circumstances where there is even
only a partial loss of
coherence of each ion, entanglement can be suddenly and completely lost \cite%
{yu06}. This has motivated us to consider the question of how
decoherence effects the scale of entanglement in the system under
consideration. Once the decoherence taken into account, i.e.,
$\gamma \neq 0$, it is very clear that the decoherence plays a usual
role in destroying the entanglement. In this case and for different
values of the decoherence parameter $\gamma ,$ we can see from
figure (\ref{dec}) that after switching on the interaction the
entanglement function increases to reach its maximum showing strong
entanglement. However its value decreases after a short period of
the interaction time to reach its minimum. The function starts to
increase its value again however with lower local maximum values
showing a strong decay as time goes on. Also, from numerical results
we note that with the increase of the parameter $\gamma $, a rapid
decay of the entanglement (entanglement sudden death) is shown
\cite{yu02,aty07}.

\begin{figure}[tbph]
\begin{center}
\includegraphics[width=7cm,height=7cm]{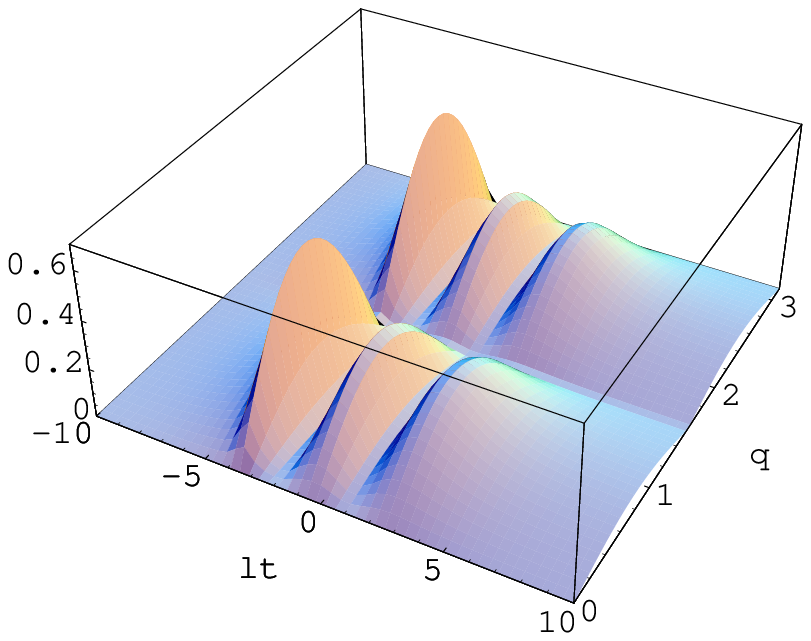} %
\includegraphics[width=6cm,height=5cm]{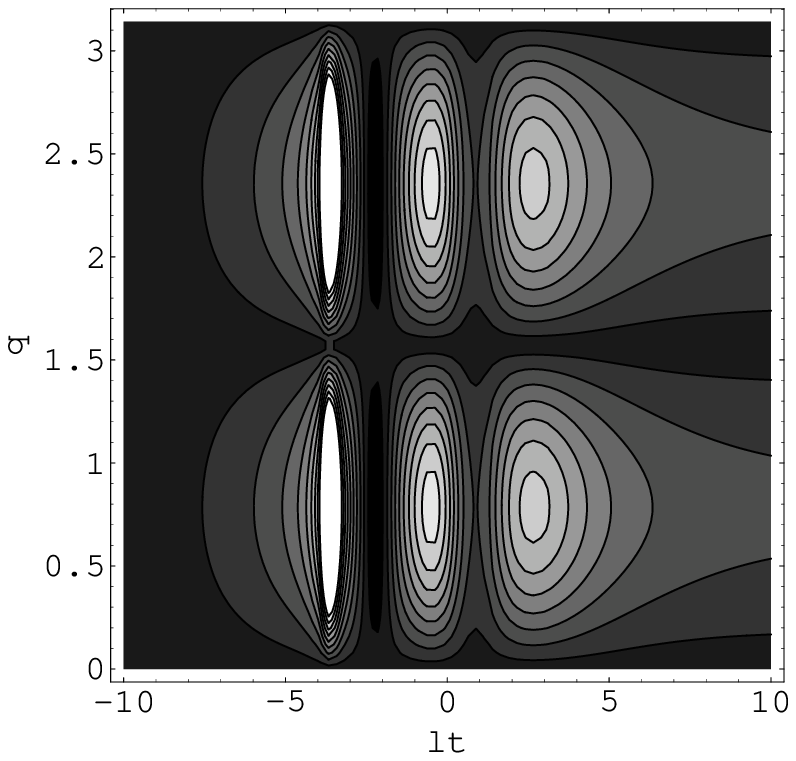}
\end{center}
\caption{The same as figure 1 but the modulated function $\protect\zeta %
(t)=\sec $h$\left( t/2\protect\tau \right) .$}
\end{figure}

The above results and connections are very intriguing, and naturally
lead us to ask what is the role played by the modulated function in
obtaining these associated phenomena of the entanglement. In order
to answer this question, we consider the modulated function $\zeta
(t)$ to be time-dependent of the form $\zeta (t)=\sec $h$\left(
t/2\tau \right) $ \cite{pra92,das99}$.$ In this form the coupling
increases from a very small value at large negative times to a peak
at time $t=0$, to decrease exponentially at large times. Thus,
depending on the value of $\tau $ and the initial time $t_{0}$,
various limits such as adiabatically or rapidly increasing (for $%
t_{0}<\lambda t\leq 0$) or decreasing (for $0\leq \lambda t<t_{0}$)
coupling can be conveniently studied. This allows us to investigate,
analytically, the effect of transients in various different limits
of the effect of switching the interaction on and off in the
ion-field system. The vanishing of the interaction at large positive
times leads to the leveling out of the inversion.

It should be noted that the time dependence specified in $\zeta (t)$
is one of a class of generalized interactions that may offer
analytical solutions. It is evident from figure 4 that the
entanglement sudden birth phenomenon is more clearly observed for
the time-dependent interaction. In this case, the entanglement
starts suddenly to build up at later times compared with the
time-independent case. At this point the entanglement from zero
evolves to its local maximum value and then oscillates with lower
local maximum followed by smooth decay (see figure 4). Although
increasing the field intensity leads to strong entanglement (maximum
value of entanglement), however the local maximum values of the
entanglement also vary and occur for some short periods of the
interaction time. This indicates that in a regime where coherent
state is considered, the underlying states are highly entangled.
Consequently, the presence of the time-dependent modulated function
increases the number of oscillations and delays the disentanglement.
All these results confirm the possibility of a practical observation
of time-dependence of the modulated function effects for prolonging
time for the disentanglement. Our results here show the important
role played by the modulated function in the entanglement dynamics
which is crucial for the onset of either entanglement sudden death
or sudden birth in the trapped ion systems considered in this paper.
In addition, our results suggest that the analytical results
presented here, could be attained for different configurations of
any two trapped ions systems \cite{sol99}.

The remaining task is to identify and compare the results presented
above for the entanglement with another accepted entanglement
measure such as the quantum relative entropy. Eisert and Plenio
\cite{eis99} have raised the question of the ordering of
entanglement measures. It has been proven that all good asymptotic
entanglement measures are either identical or fail to uniformly give
consistent orderings of density matrices \cite{vir00}. One of the
best understood cases is entanglement measure defined in terms of
the
quantum relative entropy. More explicitly, for the entangled states $\hat{%
\rho}(t)$ the quantum relative entropy is defined by the following
formula as the distance between the entangled state $\hat{\rho}(t)$
and disentangled state $tr_{\mathcal{A}}\hat{\rho}(t)\otimes
tr_{\mathcal{B}}\hat{\rho}(t)\in \mathfrak{S}(\mathcal{H}_{1}\otimes
\mathcal{H}_{2})$ \cite{fur01,lin73}
\begin{equation}
I_{\rho }\left( \rho _{t}^{\mathcal{A}},\rho _{t}^{\mathcal{B}}\right) =tr%
\hat{\rho}(t)(\log \hat{\rho}(t)-\log
(tr_{\mathcal{A}}\hat{\rho}(t)\otimes
tr_{_{\mathcal{B}}}\hat{\rho}(t))),  \label{me}
\end{equation}%
where $\rho _{t}^{\mathcal{A}}=tr_{\mathcal{B}}\left(
\hat{\rho}(t)\right) $
and $\rho _{t}^{\mathcal{B}}=tr_{\mathcal{A}}\left( \hat{\rho}(t)\right) ,$ $%
\mathcal{A(B)}$ refers to the first (second) ion$.$ Note that if the
entangled state $\hat{\rho}(t)$ is a pure state,
$S(\hat{\rho}(t))=0$ and then
$S(tr_{\mathcal{A}}\hat{\rho}(t))=S(tr_{\mathcal{B}}\hat{\rho}(t)),$
which means that we have $I_{\rho }\left( \rho _{t}^{A},\rho
_{t}^{B}\right) =2S(tr_{\mathcal{B}}\hat{\rho}(t))$. One, possibly
not very surprising, principal observation is that the numerical
calculations corresponding to the same parameters, give nearly the
same behavior with different scales. This means that the
entanglement measured by either quantum relative entropy or
concurrence measures gives rise to qualitatively the same results.
We must stress, however, that no single measure alone is enough to
quantify the entanglement in a multilevel system.

\section{Conclusions}

In summary, we have derived an intuitive extension of the standard
quantum model of two three-level trapped ions interacting with a
laser field to include the time-dependent modulated function and
intrinsic decoherence. This study reveals that the time-dependent
modulated function can be used for generating either entanglement
sudden death or sudden birth depending on a proper manipulation of
the initial state setting. We note that the existence of
entanglement sudden death reveals a fact that the non-interacting
and non-communicating two ions can abruptly lose their entanglement.
Also, it will be very interesting to extend these results to the
case of mixed states in the presence of the decoherence. We hope the
presented results can be useful for the ongoing theoretical and
experimental efforts in multi-levels particles interaction.

\bigskip

\textbf{Acknowledgment}

T. Y. acknowledges grant support from US National Science Foundation
(PHY-0758016). We are grateful to Prof. A.-S. F. Obada for helpful
discussions.
\[
\]%

\end{document}